# Optical Anisotropy in Tellurene and its Janus Allotropes - A first principle Study

Ritika Rani[1*], Munish Sharma[2] and Raman Sharma[1]

[1]*Department of Physics, Himachal Pradesh University, Shimla 171005, India.*

[2]*Department of Physics, School of Basic and Applied Sciences, Maharaja Agrasen University, Baddi 174103, India.*

(July 25, 2020)

*Corresponding Author

Email: rriti834@gmail.com

**Abstract**

Here, we present first principle study of structural, electronic, and optical properties of pristine and janus tellurene allotropes using density functional theory (DFT). The α, β, δ, and η allotrope of pristine tellurene exhibit indirect band gap while γ and σ allotropes are metallic. The bandgap shows tunability in janus tellurene compared to pristine tellurene. A metallic to semiconductor transition occurs in janus γ, σ and its allotropes. Dirac cone-like interesting feature has been observed for pristine σ tellurene which remains preserved with the opening in energy gap for janus allotrope. In optical properties, imaginary part dielectric function and electron energy loss spectra show a blue shift in janus tellurene as compared to pristine form. Static dielectric constant is tunable in janus tellurene. All allotropes of tellurene are optically active in UV-Vis region with optical anisotropy in all three directions. We believe that our findings will provide useful pointers in the experimental fabrication of devices based on janus tellurene.

# 1. Introduction

Two-dimensional (2D) materials have undergone a rapid pace of development with discoveries of alternatives to graphene such as hexagonal boron nitride (h-BN)[1], TMD's[2], phosphorene[3], Gallium Sulphide[4] etc. having inherent electronic and physiochemical properties[5]. Among various 2D materials, some materials despite having extraordinary properties are seriously impeded with disadvantages. For example, lack of bandgap in graphene which limit its direct application in devices[6], black phosphorus only conserve stability under special conditions which restrains its use under the open air[7] and many do not exhibit a direct band gap[8]. Among these 2D materials, $MoS_2$ has been identified both experimentally and theoretically as a promising material with an inherent direct band gap of 1.80 eV for various applications because its properties can be tuned by controlling the number of layers[9], electric field[10], mechanical strain[11], etc. Furthermore, it has been reported experimentally[12], [13] and theoretically[14] that TMDs based janus structures can break the inversion symmetry which results in structural anisotropy leading to interesting properties such as large piezoelectricity[12] and Rashba spin splitting[15][16].

Very recently, group VI based, a new class of mono-elemental layered structures called 'Tellurene' have been successfully synthesized [17], [18],[19] which show structural analogy to TMDs. Experimentally, tellurene has been found to exhibit high on/off ratio[20], excellent photoconductivity[21], non-linear optical response[22], high thermoelectric[23] and piezoelectric [24], [25] properties. These properties make tellurene a potential candidate for variety of applications such as FET[20], photodetectors [26], batteries[27]etc.

Various distinct layered phases of Te have been predicted such as α phase (1T-$MoS_2$ like), tetragonal β and γ (2H $MoS_2$ like) depending upon orientation of cleavage plane[17] . Recently, two more stable allotropes of tellurene have been predicted namely δ and η-tellurene[28] and a new stable phase of monolayer tellurene have been synthesized on Cd-terminated CdTe wafer surfaces[29]. Besides, many ongoing experimental and theoretical studies on tellurene, no particular attention has been given to understand of optical properties of tellurene in distinct phases. In this work, we explore structural, electronic and optical properties of six different phases of tellurene and its janus allotropes. Janus structures have been chosen as they offer an advantage of good growth control and more feasible fabrication.

Here, we explore five different allotropes of pristine and janus tellurene by using density functional theory (DFT) with the aim to study i) structural properties of pristine and janus tellurene allotropes , ii) modulation in electronic band structure, carrier effective masses, iii) tunability in optical properties including static dielectric constant, absorption spectra and electron energy loss spectra.

## 2. Simulation Details

A Vienna ab-initio Simulation Package (VASP)[30] has been used to get optimized geometry structures of 2D tellurene within the framework of density functional theory (DFT). The exchange-correlation energies were treated using generalized gradient approximation[31] of the Perdew-Burke-Ernzerhof scheme[32] . The plane wave cut-off energy of 400 eV has been used. The atomic positions were fully optimized until the forces on all atoms were less than $10^{-6}$ eV/Å. The energy convergence criteria is set to be $10^{-4}$ eV/cell. We adopt a 15 × 15 × 1 grid of k-mesh for Brillouin zone integration[33]. The mesh density of k points was kept fixed when calculating the properties for all the Te phases. In order to mimic the 2D system, we have created a vacuum space of ~20 Å along the z-direction for all phases. For optical property calculation, we have calculated the frequency-dependent dielectric matrix using k-mesh of 30x30x1 to sample the Brillouin zone finely. The total dielectric function is given by $\varepsilon(\omega) = \varepsilon_1(\omega) + i\varepsilon_2(\omega)$, where $i\varepsilon_2(\omega)$ is the imaginary part and $\varepsilon_1(\omega)$ is the real part of the dielectric function. The Krammers Kroning relations[34] were used to get real part of dielectric function and other related properties[35].

## 3. Results and Discussion

### 3.1 Structural Properties

The geometric structures for six different allotropes of two-dimensional tellurene have been depicted in figure 1. Out of six different allotropes α and γ have honeycomb-like geometry whereas β, δ, η and σ have non-honeycomb geometry. The α and γ phases have 1T and 2H $MoS_2$ like structure containing three atoms in a unit cell. The central Te atom of α and γ have larger coordination numbers as compared to atoms on top and bottom atom in a monolayer. It results in dual nature (metallic and semiconducting) similar to Mo atom in $MoS_2$ monolayer.

β-phase can be obtained from the α-phase by shifting the mid-layer Te atoms slightly towards their closest neighboring helical chains, resulting in the periodic mirror planes perpendicular to the x-direction[36], [37]. The δ-structure (figure 1d) can be obtained by artificially compressing a 2D assembly of the native α-Te helices in the direction normal to the 2D layer [28]. Another stable allotrope, labeled as η (figure 1e) can be obtained by considering a series of reflections in δ-structure [28]. Both δ-phase and η-allotrope of tellurene have six atoms per unit cell. We predict a new 2D layered structure named as σ-Tellurene (figure 1f). It has buckled structures similar to other layered materials such as silicene[38] and germanene[39] with chair-like conformation. The equilibrium lattice parameters, and other structural parameters of tellurene and its janus structures are listed in table 1. Most of our calculated values for pristine and janus structures are in good agreement with the available values in the literature [40], [41], [17], [42]. For example, for α-tellurene, our calculated lattice constant of 4.23 Å, is in excellent agreement with value of 4.24 Å reported by Wang et. al[40]. Similarly, janus $Te_2Se$ have a lattice constant of 4.10 Å which is in complete agreement with value reported by Chen et. al.[41].

Looking at modulation in bond lengths no significant change in $d_{Te-Te}$ has been observed for pristine tellurene, while a decrease in $d_{Te-X}$ bond length (where, X=Se and S) by less than 25 % have been observed in janus monolayers. These modulations (out-of-plane) in the interatomic distance leads to the change in a layer thickness (*h*). The modulation in $d_{Te-X}$ bond length can be attributed to the large(small) electronegativity of S(Se) atom. It is noteworthy that the optimization of janus monolayer does not result in any noticeable structural transition as far as the atomic arrangements are concerned. This is confirmed by calculating the structural anisotropy factor as *κ = (a − b)/(a + b)*, where *a* and *b* are lattice parameters (table 1). Note that among all considered phases α, γ, and σ exhibit isotropic nature while δ, η, and β exhibit anisotropic nature. The structural anisotropy is maximum in δ phase and minimum in β phase. A negligible change in *κ* can be seen in janus monolayer compared to pristine monolayer.

To ensure the stability of janus monolayer we have calculated the formation energy per atom for α, β, γ, δ and η as:

$$E_f = \frac{1}{3m}\left[E_{m(Te_n X_{n-1})} - \left(E_{m(Te_n)} + E_{m(X_{n-1})}\right)\right]$$

and for σ as:

$$E_f = \frac{1}{2m}\left[E_{Te_{n-1}X_{n-1}} - \left(E_{Te_{n-1}} + E_{X_{n-1}}\right)\right]$$

Here $n=2$ and $m=1$ for α, β, γ & σ and $m=2$ for δ, and η. In the above equations first-term represents the total energy of the composite system and the second & third terms represent total energy of isolated Te and X (=Te, Se, S) atoms, respectively. Obtained formation energy has been tabulated in table 1. Negative value of $E_f$ indicate the exothermic nature of formation. Formation energy is more negative for janus structures as compared to a pristine monolayer in respective phases (β, δ, η, and σ) indicating favorability of janus structures. This is attributed to the structural asymmetry (different bond lengths) present in these allotropes offering favorability of formation. In α and γ phase, the formation of janus structure are less favorable because of the presence of high structural symmetry which leads to a strong covalent bond.

**Table 1** Calculated values of equilibrium lattice parameters (*a*, *b*), bond length (*d*), thickness (*h*), anisotropy factor (*κ*) and formation energy (*E_f*) for Janus tellurene allotropes using GGA-PBE functional. Notation IP, OP represents in-plane atomic bond length and out of plane bond lengths respectively.

| Phase | X | Lattice Constant | Bond Length (d) | | Thickness (h) | κ | $E_f$ |
|---|---|---|---|---|---|---|---|
| | | | $d_{Te-Te}$ | $d_{Te-X}$ | | | |
| α | Te | a=b=4.23,4.24[a], 4.45[b],4.15[c] | 3.04, 3.02[a],3.02[c] | - | 3.62 | - | -2.88 |
| | Se | 4.10, 4.02[d], 4.10[e] | 3.02, 3.14[d] | 2.85, 2.84[e] | 3.46, 3.41[d] 3.45[e] | | -3.59, -2.01[d] |
| | S | 4.02 | 3.00 | 2.72 | 3.32 | - | -4.17 |
| β | Te | a = 5.68, 5.69[a],4.17[c] b = 4.22, 4.23[a],5.47[c] | 4.44, 3.03 | - | 2.77 | 0.147 | -3.23 |
| | Se | a = 5.45, b = 4.10 | 4.18 | 2.88 | 2.57 | 0.141 | -4.62 |
| | S | a = 5.30, b = 4.09 | 3.93 | 2.61 | 1.81 | 0.128 | -4.26 |
| γ | Te | a = b = 3.96, 3.92[f] | 3.08, 3.08[f] | - | 4.12 | - | -3.31 |
| | Se | a = b = 3.93 | 3.05 | 2.90 | 3.85 | - | -2.60 |
| | S | a = b = 3.87 | 3.04 | 2.80 | 3.72 | - | -3.14 |
| δ | Te | a = 10.17, b = 4.23 | 2.64, 2.48 (IP) | - | 2.53 | 0.412 | -5.86 |

| | | | 2.59 (OP) | | | |
|---|---|---|---|---|---|---|
| | Se | a = 9.81, b = 4.10 | 3.01 | 2.93 (IP) 2.57 (OP) | 2.47 | 0.410 | -7.42 |
| | S | a = 9.33 b = 4.22 | 2.54 | 2.81 (IP) 2.42 (OP) | 2.35 | 0.377 | -8.77 |
| η | Te | a = 8.56, b = 4.20 | 3.03, 2.79 (IP) 3.01 (OP) | - | 4.08 | 0.341 | -6.66 |
| | Se | a = 8.37, b = 4.08 | 2.98 | 2.59 (IP) 2.87 (OP) | 4.01 | 0.342 | -7.70 |
| | S | a = 8.22, b = 4.03 | 2.98 | 2.44 (IP) 2.72 (OP) | 3.81 | 0.344 | -8.75 |
| σ | Te | a = b= 4.10, 4.08[g] | 3.20, 3.03 | - | 0.90 | - | -3.36 |
| | Se | a = b = 3.88 | - | 3.11, 2.87 | 0.83 | - | -3.57 |
| | S | a = b = 3.74 | - | 3.04, 2.75 | 0.75 | - | -3.85 |

[a]Ref [40]  [b]Ref[42]  [c]Ref[43] [d]Ref[44]  [e]Ref[41]  [f]Ref[17]  [g]Ref[45]

Furthermore, to know more about change in structural properties we have examined the charge density redistribution in different allotropes of tellurene (figure not shown here). The charge density difference is calculated by taking the difference between the total charge density of the composite system of janus tellurene monolayer and the sum of isolated charge density of Te and X atoms. In pristine tellurene charge accumulation occurs in the vicinity of covalent bond while in janus case most of the charge accumulates around selenium and sulfur atom. This type of behavior is obvious due to the high electron affinity of X atoms as compared to tellurium. The charge redistribution has been found same along two planar directions in α, γ and σ allotropes while all other phases have different charge redistribution along two directions. For example, the electrons are symmetrically localized along both armchair and zigzag directions in α and γ allotropes while charge redistribution is not symmetric in other cases. This asymmetric charge redistribution in β, δ, and η allotropes confirms the presence of structural anisotropy in these phases.

**3.2 Electronic Properties**

The band structures of tellurene with different allotropes have been shown in Figure 2. Most of the allotropes of monolayer tellurene are predicted to exhibit a broad range of indirect band gaps semiconductors[17], [28], [40], [41], [45]. Our calculated values of band gaps (table 2) are in very good agreement with the available values in the literature[17], [28], [40], [41], [45] e.g., indirect band gaps of α-Te (0.66 eV) and β-Te ( 1.25 eV) for pristine case are in very good agreement with the previously reported values of (0.65 eV) and (1.09 eV)[40], respectively. Similarly, for janus tellurene (α-$Te_2$Se), we find a bandgap of 0.75 eV which is in close agreement with earlier reported DFT-PBE based value (0.73eV)[41]. Additionally, we find a tunability of band gap in janus tellurene allotropes as compared to pristine tellurene allotropes. All the janus allotropes have generally higher bandgap as compared to pristine form except for δ allotrope. Interestingly, a metallic to semiconductor transition has been found in janus σ and γ allotrope of tellurene. In η-Te for pristine case we find indirect band gap of 0.56 eV, however previous results by D. Liu et. al. [28] predict η-Te to be direct band gap 0.3 eV. The mismatch between two studies arises due to missing crucial Brillion zone direction in Liu et.al.[28] study along which band gap lies. A conical shaped linear dispersion along Γ-$M_1$ direction make σ interesting phase with expected massless charge carriers at VBM and CBM. It is worth mentioning here that conical dispersion in σ allotrope is different from the Dirac dispersion observed in graphene. The Dirac dispersion in graphene is at the zone centre while in janus σ tellurene allotrope dirac like dispersion is along Γ-$M_1$ direction (towards Brillouin zone boundary)[46]. Note that the Dirac cone-like feature remains preserved in janus σ allotrope. To summarize, band gap opening occurs in janus allotropes of tellurene which lies in mid-infrared (0.3 eV) to near-infrared (1.5 eV) spectral range.

Table 2. Calculated electronic band gap ($E_g$, in eV), carrier effective masses (m*) of electrons ($m_e$) and holes ($m_h$) at band edges, work function (Φ, in eV) and deformation potential ($E^d$) for valance band ($E_V^d$) and conduction band ($E_C^d$).

| Allotrope | X | $E_g$ | $m_h^*(m)$ | $m_e^*(m)$ | Φ | $E^d$ | |
|---|---|---|---|---|---|---|---|
| | | | | | | $E_V^d$ | $E_C^d$ |
| α | Te | 0.66, 0.65[a], 0.92[b], 0.94[exp], 0.76[c] | 0.01 | 0.002 | 4.69, 4.67[i] | - | - |
| | Se | 0.75, 0.72[d] 0.72[e] | 0.02 | 0.002 | 4.96 | -0.06 | 0.03 |
| | S | 1.00 | 0.14 | 0.002 | 5.39 | -0.19 | 0.15 |
| β | Te | 1.25, 1.09[a], 1.17[c] | 0.004 | 0.020 | 4.33 | - | - |
| | Se | 1.38 | 0.004 | 0.017 | 4.62 | 0.00 | 0.13 |
| | S | 1.47 | 0.004 | 0.016 | 4.77 | -0.01 | 0.21 |
| γ | Te | 0.00, 0.00[f] | - | - | 4.57 | - | - |
| | Se | 0.36 | 0.015 | 0.018 | 4.82 | -0.18 | 0.18 |
| | S | 0.50 | 0.016 | 0.025 | 4.93 | -0.25 | 0.25 |
| δ | Te | 1.70, 0.9[g] | 0.007 | 0.006 | 4.10 | - | - |
| | Se | 1.21 | 0.017 | 0.007 | 4.56 | 0.03 | -0.46 |
| | S | 1.65 | 0.007 | 0.009 | 4.77 | 0.01 | -0.04 |
| η | Te | 0.56, 0.3[g] | 0.005 | 0.005 | 4.24 | - | - |
| | Se | 0.66 | 0.018 | 0.005 | 4.33 | -0.05 | 0.05 |
| | S | 0.85 | 0.005 | 0.075 | 4.39 | -0.15 | 0.14 |
| σ | Te | 0.00, 0.00[h] | - | - | 4.44 | - | - |
| | Se | 0.21 | 0.0004 | 0.009 | 4.83 | -0.07 | 0.14 |
| | S | 0.25 | 0.0007 | 0.005 | 4.93 | -0.13 | 0.12 |

[a]Ref[40]　　　[b]Ref[42]　　　[c]Ref[43]　　　[d]Ref[44]　　　[e]Ref[41]
[f]Ref[17]　　　[g]Ref[28]　　　[h]Ref[45]　　　[i]Ref[47]　[exp]Ref[48]

To understand the contribution to the energy bands in band structure we have analyzed atom projected density of states (PDOS) as shown in Figure 3. The projected density of state reveal that the states near the Fermi level are mainly contributed by *Te-p* orbitals in pristine allotropes. In janus α, β, and γ tellurene the VBM is composed of strongly hybridized *Te-p* and *X-p* states (*X* = *S* and *Se*), while CBM is mainly composed of *Te-p* states. For δ-Te, η-Te, both VBM and CBM are dominated by *Te-p* state. In janus σ tellurene a considerable contribution of *X* atom has been found in the vicinity of Fermi energy

Furthermore, knowledge of the carrier effective mass, work function, and electronic band deformation potentials are important to account for transport phenomenon, device performance, and device modeling. All these parameters can be calculated from the electronic band structure. The calculated carrier effective mass, work function, and electronic band deformation potentials have been tabulated in table 3. The modulations in electronic band structures can be quantified as deformation potentials. The modulation in the energy of valance band and conduction band is quantified as valance band deformation potential ( $E_V^d$ ) and conduction band deformation potential ( $E_C^d$ ).

Experimentally, the STM and µ-XPS techniques can be used to measure these deformations[49]. We define the valance band deformation ( $E_V^d$ ) as $E_V^d = E_V^{pristine} - E_V^{Janus}$ and conduction band deformation ( $E_V^d$ ) as $E_C^d = E_C^{pristine} - E_C^{Janus}$. Calculated valance band and conduction band deformations have been tabulated in table 2. The negative values indicate the upward shift and positive values as downward shift in VBM and CBM respectively. This shift in VBM and CBM is principal cause of band gap opening in janus tellurene allotropes.

The carriers effective mass of electron ( $m_e^*$ ) and hole ( $m_h^*$ ) were determined from the curvature of the energy band at conduction band minimum. (CBM) and valence band maximum (VBM) respectively. The $m_e^*$ and $m_h^*$ can be calculated as:

$$m^* = \frac{\hbar^2}{\frac{\partial^2 E}{\partial k^2}}$$

where *E* and *k* correspond to the energy and the reciprocal lattice vector. The term in denominator represent the slope of energy bands at VBM and CBM. The effective mass has been found to be maximum for β phase and minimum for α phase in pristine form. Note that our calculated values of effective mass for pristine allotropes are generally less than 2D TMDs ($MoX_2$; where X = S, Se and )[50]  but greater than that of graphene, β-sillicine and β-garmanene[4]. In general, the effective mass of holes are found to be less than that of electron in pristine tellurene indicating dominance of holes as charge carriers in pristine tellurene. Our inferences drawn here resembles with the earlier studies for the β phase [17], [37], [51].

Note that the hole effective mass increases considerably while negligible modulation in electron effective mass have been observed in janus tellurene as pristine case in all phases. The increase in hole effective mass in janus tellurene is attributed to the change in VBM due to structural modulation (bond length,) which in principle leads to different hybridization of atomic orbital and hence change in the electronic band structure. Hence, janus tellurene offers tunability in carrier effective mass (holes effective mass) and deformation potentials which leads to expected tunability in holes mobilities in janus tellurene based devices. Furthermore, work function ($\Phi$) of the 2D materials for pristine and Janus case for different phase has been calculated as $\Phi = E_{vac} - E_f$ ; where $E_{vac}$ and $E_f$ represent the vacuum potential and Fermi energy, respectively. The work function for Te monolayer for α-Te (pristine) is 4.69 eV which is less than that of MoS$_2$ monolayer (5.51 eV)[50] and quite resemble with the recently calculated value of 4.67 eV reported by J. Singh *et. al.*[47]. An increase in work function has been found for janus allotropes as compared to pristine allotropes. The increase in work function for janus allotropes is attributed to modulation in electrostatic potential and hence an increase in work function.

### 3.3 Optical Properties

Tellurene has been observed to be a potential candidate for applications in optoelectronic devices due to the presence of *Te-p* orbital in valance band and conduction band leading to novel optical response as compared to *s-p* hybridized materials[34]. Next, we explore the optical properties of pristine tellurene and janus tellurene in six different phases. The optical properties have been calculated for electric field vector parallel (E ∥ c) and perpendicular (E ⊥ c) to the c-axis. The imaginary part of dielectric function ($\varepsilon_2$) is evaluated using the following relation [52].

$$\varepsilon_2^{\alpha\beta} = \frac{4\pi^2 e^2}{\Omega} \lim \frac{1}{q^2} = \sum_{c,v,k} 2\omega_k \delta(\varepsilon_{ck} - \varepsilon_{vk} - \omega) u_{ck} + e_\alpha q | u_{vk} u_{ck} + e_\beta q | u_{vk}$$

here, superscript α and β represent cartesian components, v and c represent valance and conduction band and k represents k-points of the Brillouin zone; $e_\alpha$ and $e_\beta$ are unit vectors along x, y, and z-direction; $\varepsilon_{ck}$ and $\varepsilon_{vk}$ represent the energy of valence and conduction band and $u_{vk}$ and $u_{uk}$ refers to cell periodic part of orbitals. Hence real part ($\varepsilon_1$) can be obtained easily if we calculate the imaginary part firstly ($\varepsilon_2$). The anisotropic complex dielectric function for in-plane polarization (E⊥c) and out-of-plane polarization is obtained using the following relations[34] $\varepsilon_x^\perp(\omega) = \varepsilon^{xx}(\omega)$, $\varepsilon_y^\perp(\omega) = \varepsilon^{yy}(\omega)$ and $\varepsilon_z^\perp(\omega) = \varepsilon^{zz}(\omega)$, where $\varepsilon^{xx}(\omega)$, $\varepsilon^{yy}(\omega)$ and $\varepsilon^{zz}(\omega)$ are diagonal elements of the dielectric matrix.

Figure 4 and Figure 5 show imaginary part of dielectric function (ε₂) in-plane polarization (E⊥c) and out-of-plane polarization (E ∥ c) for pristine and janus tellurene allotropes, respectively. The corresponding peak positions have been tabulated in table 3 which correspond to inter-band transitions in the electronic band structure. Note that the excitonic effects can be studied by using computationally expensive and advanced method such as solving Bethe Salpeter equation. Pristine tellurene exhibit broad structure peak in Visible region for α, γ, σ allotropes and in UV region for β, δ, and η allotropes when E∥c. A blue shift in structural peak has been observed for janus tellurene not observed in pristine. For in-plane polarization (E⊥c) pristine tellurene shows prominent structure peak in visible region. A strong out of plane optical anisotropy has been found for β, δ, and η tellurene while imaginary part of dielectric function show isotropic nature for α, γ, [50] and σ allotropes which get blue shifted for janus tellurene. For example, α tellurene shows isotropic structure peak at 1.41 eV which get shifted by ~ 0.33 eV for janus tellurene. On the other hand, β tellurene shows anisotropic structure peak at 1.75 eV and 1.97 eV with a considerable blue shift for janus tellurene. The anisotropic nature was also predicted by Sang *et. al*. for β tellurene[34]. It is worth mentioning here that the broadened peak in the imaginary part of dielectric function $\varepsilon_2$ correspond to a large number of inter-band transitions due to the presence of different number of energy bands along different Brillouin zone directions.

Table 3 Peak positions in imaginary part of dielectric function $\varepsilon_2$ for pristine and janus tellurene.

| Polarization | $E \parallel c$ | | | $E \perp c$ | | | | | |
|---|---|---|---|---|---|---|---|---|---|
| X | Te | Se | S | Te | | Se | | S | |
| System | | | | $E_x$ | $E_y$ | $E_x$ | $E_y$ | $E_x$ | $E_y$ |
| α | 2.83 | 2.74 | 2.77 | 1.41 | 1.41 | 1.49 | 1.49 | 2.27 | 2.27 |
| β | 5.47, 5.00[a] | 5.85 | 6.40 | 1.75 | 1.97, 2.40 | 1.87 | 1.87, 2.81 | 2.13 | 2.37, 2.84 |
| γ | 2.37 | 2.75 | 2.79 | 0.95 | 0.95, 1.90 | 1.00 | 1.00 | 1.27 | 1.77 |
| δ | 5.61 | 5.31 | 6.24 | 2.28 | 3.33, 2.49 | 1.77 | 1.77 | 2.21 | 2.01 |
| η | 2.73, 3.46 | 3.72 | 4.00 | 2.37 | 2.19 | 2.54 | 2.54 | 1.80 | 2.60 |
| σ | 7.55 | 8.04 | 8.54 | 0.48 | 0.48 | 0.80 | 0.80 | 0.82 | 0.82 |

[a]Ref[34]

Furthermore, frequency-dependent real part of the dielectric function $\varepsilon_1$ has been plotted in Figure 6 and Figure 7. The calculated values of real part of a dielectric function at zero frequency can be quantified as static dielectric constant $\varepsilon_s$ which are tabulated in table 4. The static dielectric constants are calculated by the contribution from optical transitions taking place in entire frequency range. Note that the context of zero frequency here should be unambiguously understood as frequency lower than interband transition frequency but greater than photon frequency[11]. The pristine and janus tellurene show modulation in value of static dielectric constant for both polarizations. The modulations in $\varepsilon_s$ exhibit inverse relationship with electronic bandgap which is consistent with Penn model[54]

$$\varepsilon_2(0) \approx 1 + \left(\frac{\omega_p}{E_g}\right)^2$$

The characteristic frequency value where $\varepsilon_1$ crosses the zero axis from negative side indicates the plasmon frequency and it helps to understand the dielectric behavior of material. The observed plasmon frequency has been tabulated in table 5. We find the presence of plasmon in all allotropes of pristine and janus tellurene. The plasmon frequency is blue-shifted for janus tellurene as compared to pristine tellurene in all phases. At high frequency greater than ~6 eV all materials behave like dielectrics when E⊥c. For E∥c, tellurene in α, γ and η become dielectric at a lower frequency (~7 eV) while β, δ, and σ behaves like dielectric above ~10 eV. A similar behavior has been observed for janus tellurene with slight modulations.

**Table 4** Static dielectric constant at for pristine and janus tellurene when E ∥ c and E⊥c.

| Polarization | *E* ∥ *c* | | | *E* ⊥ *c* | | | | | |
|---|---|---|---|---|---|---|---|---|---|
| X | Te | Se | S | Te | | Se | | S | |
| System | | | | $E_x$ | $E_y$ | $E_x$ | $E_y$ | $E_x$ | $E_y$ |
| α | 6.82 | 6.12 | 5.52 | 14.46, 13.9[a] | 14.44 | 12 | 12 | 10.47 | 10.45 |
| β | 2.13 | 2.01 | 1.99 | 7.49 | 6.36, 5.9[a] | 6.51 | 5.87 | 5.78 | 5.66 |
| γ | 6.07 | 5.40 | 4.94 | 12.72 | 12.91 | 8.85 | 8.83 | 7.85 | 7.79 |
| δ | 1.52 | 1.47 | 1.38 | 2.60 | 2.54 | 3.69 | 4.44 | 2.79 | 2.93 |
| η | 2.36 | 2.28 | 2.90 | 3.16 | 4.12 | 3.16 | 3.96 | 2.95 | 3.60 |
| σ | 1.63 | 1.56 | 1.52 | 64.18 | 74.61 | 27.89 | 28.56 | 23.66 | 23.80 |

[a]Ref[47]

**Table 5** Plasmon frequency ($\omega_p$) for pristine and Janus tellurene when E || c and E⊥c.

| Polarization | *E* \|\| *c* | | | *E* ⊥ *c* | | | | | |
|---|---|---|---|---|---|---|---|---|---|
| X | Te | Se | S | Te | | Se | | S | |
| System | | | | $E_x$ | $E_y$ | $E_x$ | $E_y$ | $E_x$ | $E_y$ |
| α | 6.15 | 6.48 | 6.82 | 6.38 | 6.38 | 6.73 | 6.73 | 7.07 | 7.07 |
| β | 6.78 | 7.02 | 7.83 | 5.69 | 6.56 | 6.09 | 7.02 | 7.12 | 7.83 |
| γ | 4.98 | 5.01 | 7.62 | 4.99 | 4.99 | 5.26 | 5.26 | 6.35 | 6.35 |
| δ | 7.28 | 6.49 | - | 4.16 | 4.16 | 5.11 | 4.92 | 3.62 | 3.82 |
| η | 4.56 | 4.90 | 5.00 | 2.73 | 4.74 | 3.33 | 5.09 | 3.40 | 5.20 |
| σ | 8.77, 9.02 | 9.38 | 9.64 | 1.46, 6.58 | 1.46, 6.58 | 1.60, 7.24 | 1.60, 7.24 | 1.80, 7.71 | 1.80, 7.71 |

The electron energy loss function has been calculated by using $\varepsilon_1$ and $\varepsilon_2$ (real and imaginary dielectric constant). Electron energy loss (EEL) function is inversely proportional to square of dielectric functions and corresponds to the collective excitations of electrons of system. It has been determined from the following expression:

$$-\operatorname{Im}\left\{\frac{1}{\varepsilon(\omega)}\right\} = \frac{\varepsilon_2(\omega)}{\varepsilon_1^2(\omega) + \varepsilon_2^2(\omega)}$$

Calculated EEL spectra has been reported in Figure 8 and Figure 9 and peak positions have been tabulated in table 6. The α, β,and γ show two intense plasmon peak between 5-10 eV while a single peak for δ, η, and σ allotropes when light is out of plane polarized. Janus tellurene show an observable blue shift not seen in pristine tellurene for E||c although the shift is not uniform. EEL peaks are broader when light is polarized along x-axis and sharp along y-axis. A shift of about 0.34 eV and 0.60 eV have been observed for Se and S containing janus tellurene as compared to pristine tellurene. To summarize, EEL spectra exhibit highly anisotropic nature as far as the light polarization is concerned. Janus tellurene shows similar trends with a considerable blue shift.

**Table 6** Electron energy loss spectra for pristine and janus tellurene when E ∥ c and E⊥c.

| Polarization | $E \parallel c$ | | | $E \perp c$ | | | | | |
|---|---|---|---|---|---|---|---|---|---|
| X | Te | Se | S | Te | | Se | | S | |
| System | | | | $E_x$ | $E_y$ | $E_x$ | $E_y$ | $E_x$ | $E_y$ |
| α | 6.38, 6.20[a], 9.22, 9.0[a] | 6.73, 8.72, 9.47 | 7.32, 8.84, 9.59 | 6.38 | 6.38 | 6.98 | 6.98 | 7.32 | 7.32 |
| β | 7.00 | 7.48 | 7.83 | 5.91 | 6.56 | 6.09 | 7.26 | 7.12 | 7.83 |
| γ | 4.98, 7.36, 8.07 | 5.01, 7.72 | 5.08, 7.87 | 4.98 | 4.98 | 7.02 | 7.02 | 7.11 | 7.11 |
| δ | 7.49 | 6.49 | 6.85 | 4.37 | 4.37 | 5.11 | 4.92 | 4.03 | 5.23 |
| η | 4.56 | 4.90 | 5.20 | 3.28, 6.07 | 4.92 | 3.33 | 5.29 | 3.56 | 5.40 |
| σ | 9.02 | 9.65 | 10.19 | 5.36, 6.87, 8.53 | 8.77 | 7.51, 9.12 | 7.51, 9.12 | 7.99, 8.54, 9.37 | 7.99, 8.54, 9.37 |

[a]Ref[47]

## 4. Conclusions

In summary, we have used density functional theory (DFT) to obtain structural, electronic and optical properties of different allotropes of tellurene. Structural modulations in terms of $d_{Te-X}$ bond length (X = Se and S) has been found for janus tellurene. Pristine tellurene has been found to be semiconducting in α, β, δ, and η phase and conducting in γ and σ phase. Semiconducting allotropes of pristine tellurene exhibit band gap ranging between 0.56-1.75 which can be further tuned in janus allotropes. A metallic to semiconducting transition has been observed in janus γ and σ allotropes. A conical shaped Dirac cone-like linear dispersion in σ allotrope makes it an interesting allotrope for further studies. Effective mass of holes in tellurene has been found smaller than electrons which indicates higher carrier mobility of holes over electrons. The effective mass of holes is less than that of electrons also in many other 2D materials like germanane, h-BN, SiC and graphane etc. Imaginary part of dielectric function shows a blue shift in both perpendicular and parallel polarization for janus tellurene as compared to pristine tellurene. A similar blue shift has been observed in EEL spectra. Static dielectric constant for janus allotropes are lower than pristine allotropes. The β, δ, and η allotropes exhibit optical anisotropy in both perpendicular and parallel polarization. Tellurene exhibit 3D anisotropy with stronger absorbance in the UV region which can be highly beneficial for developing polarized optical sensors.

**Conflict of Interest**

There is no conflict of interest to declare.

**Acknowledgements**

To obtain the results presented in this paper we have used CVRAMAN high performance computing facility at Physics Department Himachal Pradesh University. Ritika wish to thank Himachal Pradesh University for providing financial support in the form of HPU JRF.

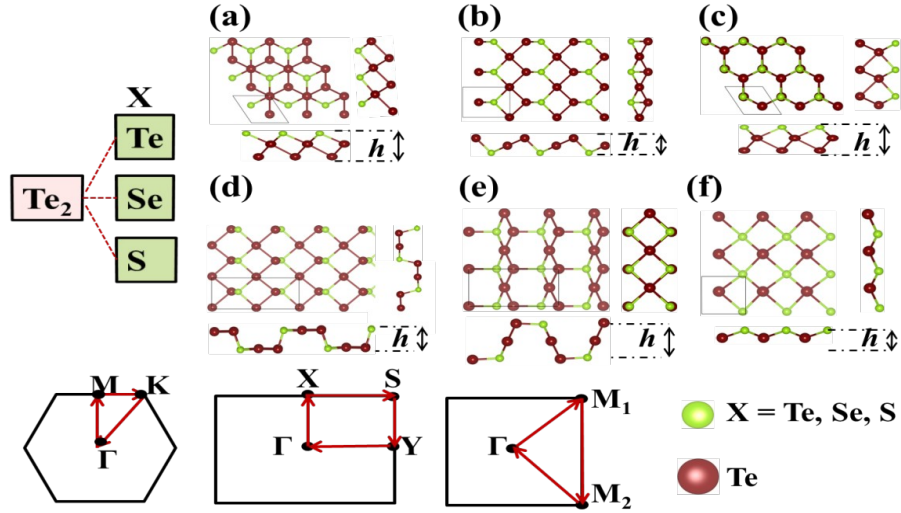

**Figure 1.** Ball and stick model of two dimensional tellurene allotropes. Red color balls represents tellurium and green color represents X atoms (X = Te, Se and S). The brillouin zone for these allotropes are also given. The unit cell is marked by a blue color. Parameter $h$ represent the thickness of monolayer.

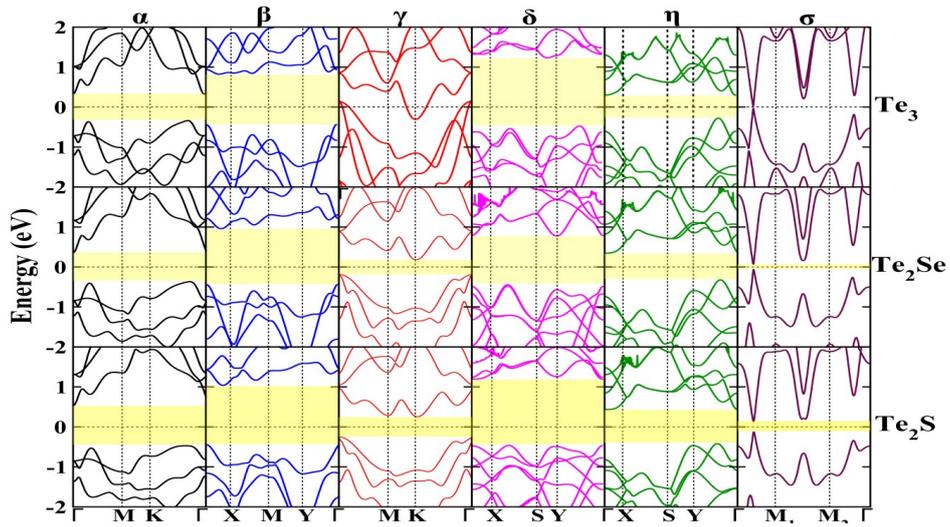

**Figure 2** Electronic band structure for six different allotropes of pristine and janus tellurene monolayer using PBE functional. Shaded region indicate the band gap. Fermi level set at 0 eV.

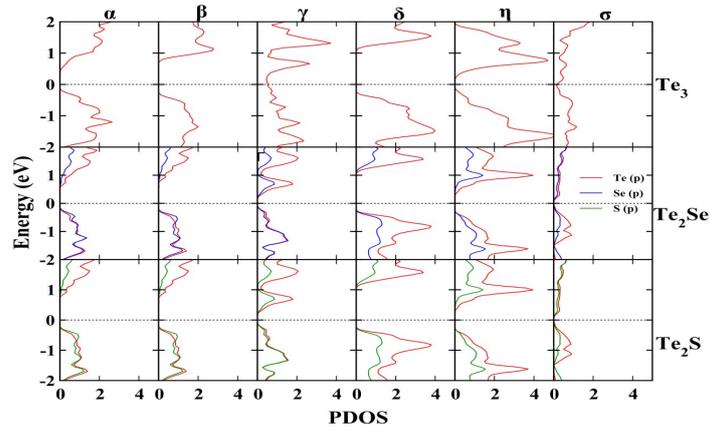

**Figure 3.** Atom projected density of states for different allotropes of tellurene and their janus structures.

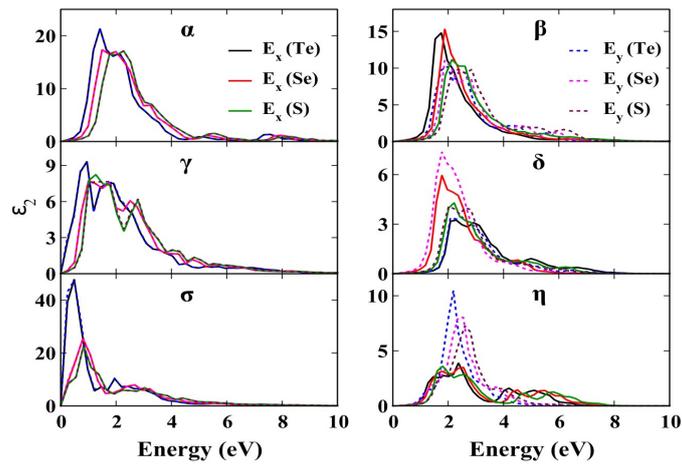

**Figure 4.** Imaginary part of dielectric function for E⊥C for different allotropes of pristine and janus tellurene.

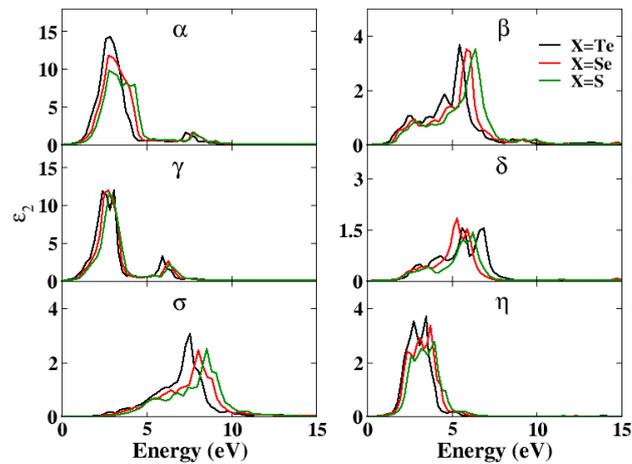

**Figure 5** Calculated imaginary part of dielectric function for E∥C for different allotropes of pristine and janus tellurene.

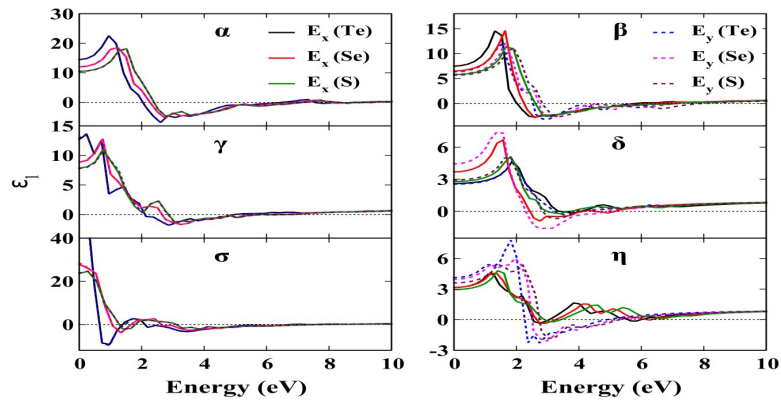

**Figure 6** Calculated real part of dielectric function for E⊥C for different allotropes of pristine and janus tellurene.

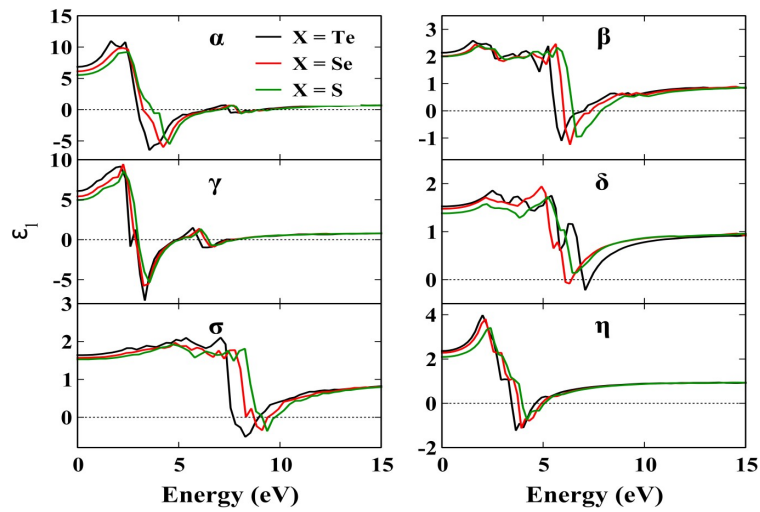

**Figure 7** Calculated real part of dielectric function for E∥C for different allotropes of pristine and janus tellurene.

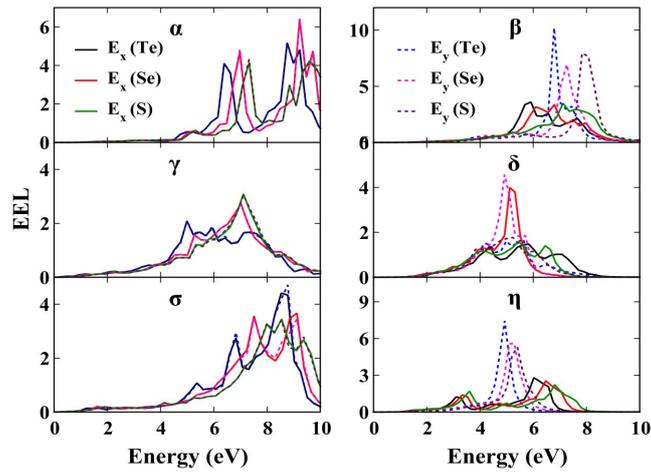

**Figure 8** Calculated electron energy loss spectra (EEL) for E⊥C for different allotropes of pristine and janus tellurene.

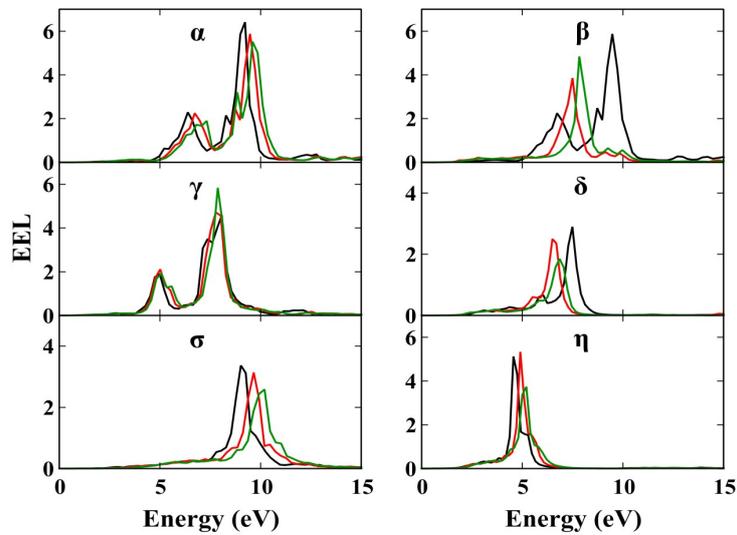

**Figure 9** Calculated electron energy loss spectra (EEL) for E||C for different allotropes of pristine and janus tellurene

25